\documentclass[11pt]{article}
\usepackage{fullpage}
\usepackage[small,compact]{titlesec}
\usepackage{graphicx,psfrag,amsmath}
\usepackage{amssymb,amsfonts,amsthm,latexsym}
\everymath{\displaystyle}
\allowdisplaybreaks
\begin{document}
\begin{center}
%\vspace*{3cm}

{\Large \bf  Symmetry structure and solution of evolution-type equations with time dependent parameters in financial Mathematics}
%\vspace{2cm}

{\large \sc M O Okelola$^{a*}$ and K S Govinder$^{a}$}
%\vspace{1cm}

{\it ${}^a$School of Mathematics, Statistics and Computer Science,\\ University of
KwaZulu--Natal,\\ Private Bag X54001, Durban 4000,\\ South Africa.\\

%${}^b$Center for Computational Finance and Economic Agents,\\
%University of Essex,\\
%Wivenhoe Park, Colchester,\\
%CO4 3SQ, United Kingdom.
}

$^*$Email: okelola@ukzn.ac.za
\end{center}

\begin{abstract}
Mathematical models with time dependent parameters are of great interest in financial Mathematics because they capture real life scenarios in the financial market. In this study, via the Lie group technique, we analyse evolution-type equations with time dependent parameters and give the general symmetry structure of these equations. In addition, we illustrate this method by looking at an example of exotic options called the power options. Our model parameters are time dependent and the option gives a continuous yield dividend at different time intervals. We present new solutions, satisfying the boundary conditions, to this important problem.
\end{abstract}


\section{Introduction}
Mathematical models  of financial derivatives were arguably pioneered by Black and Scholes (B-S) \cite{BS73} and this led to the scientific legitimization of the options market. Many empirical tests have since shown that the price from this model is fairly close to the observed price \cite{BAA08}. A large number of financial derivatives are now modeled after the B-S equation with  general form
\begin{equation}
\label{aa}
V_t + X S^{2} V_{SS} +   Y S V_{S} - Z V = 0,
\end{equation}
where $S$ is the underlying asset price, $V(S,t)$ is the option price and $X, Y, Z$ include important financial parameters like the stock volatility and interest rate. In reality though, these parameters change over time throughout the duration of the option. This necessitated the need to study equations that capture a scenario wherein the parameters are time dependent, viz.
\begin{equation}
\label{a}
V_t + X(t) S^{2} V_{SS} +   Y(t) S V_{S} - Z(t) V = 0.
\end{equation}
 In our study we give the general symmetry structure of these financial derivatives (particularly when the model parameters are time dependent) and the route to their solution satisfying accompanying boundary conditions. We illustrate this method by looking at an exotic option called the power options.
   
Exotic options are financial derivatives that have more complex features than commonly traded products. They are used to capture the increasing complexities of the financial market -- especially the nonlinear dependence of the option payoff on the asset price -- and are mostly traded over the counter to suit client needs.  Power options are generally characterized by the nonlinear payoff structure of the underlying asset price raised to a certain power ($\beta$) at expiration time. Previous studies by Cox and Rubenstein \cite{CR78} looked at the case of squared power options and Tompkins \cite{TRG99} subsequently produced the complete solution to the problem of the power parameter as any given integer value. Esser \cite{ESS04} produced solutions to the problem when the asset price is raised to any real valued number. Okelola {\it et al} \cite{OGO14} then looked at the realistic case of the model parameters being time dependent.

We note that the  PDE (\ref{a}) is linear and a study by Johnpillai and Mahomed \cite{JM01} has found criteria for parabolic linear homogeneous equations to be equivalent to the classical heat equation. They showed that if the relative invariant of the equation is zero, it can then be concluded that the equation is equivalent to the heat equation. It should however be noted that for the problem we look at in this paper, it will not be a straightforward matter to transform the solutions  and determine the {\it behavior} of equation (\ref{a}) from that of the heat equation. In addition, the solution to equation (\ref{a}) is required to satisfy certain boundary conditions and this demand cannot be forced on the heat equation in a direct manner. Due to these reasons, we decided to examine the PDE {\it apriori}.
\section{Lie group approach} The Lie group technique utilizes the notion of symmetry to obtain solutions to DEs in an algorithmic manner. Consider an  $n$th order system of DEs
\begin{equation}\label{6}
\Xi_d(t,y,\partial y,\ldots ,\partial^{n}y)=0, \hspace{2cm} d=1,\ldots,m
\end{equation}
where $t = t^1,\ldots,t^e$ are the independent variables, $y=y^1,\ldots,y^f$ are the dependent variables and $\partial^l$ is the $l$-th partial derivative
 of $y$ with respect to $t$ (for $l = 1,\ldots,n$). An  infinitesimal generator
\begin{equation}\label{7}
X = \zeta^i(t,y){\partial_{t^i}} + \eta^\mu(t,y){\partial_{y^\mu}},
\end{equation}
of the one-parameter Lie group of transformations 
\begin{subequations}\label{8}
\begin{align}
(t^*)^i = f^i(t,y;\epsilon),\\
(y^*)^\mu = g^\mu(t,y;\epsilon),
\end{align}
\end{subequations}
is said to be point symmetry of equation (\ref{6}) if and only if 
\begin{equation}\label{9}
X^{(n)} \Xi_d(t,y,\partial y,\ldots ,\partial^{n}y)=0,
\end{equation}
where $X^{(n)}$ is the $n^{th}$ prolongation of $X$ \cite{OP86}.

A local group of transformations $Z$ (acting on $M \subset T \times Y = \mathbb{R}^e \times \mathbb{R}^f$) is said to be a symmetry group of $\Xi$ if every element $z \in Z$ transforms solutions of $\Xi$ to other solutions of $\Xi$. On reduction of $\Xi$ via $Z$, every solution of the new system $(\Xi/Z)_d(a,b,\partial b,\ldots,\partial^n b)$ gives rise to a $Z$-invariant solution to $\Xi$. This technique has been used to great effect in Financial Mathematics \cite{COG10, LOS06}.
\section{Symmetry structure of the problem} The PDE (\ref{a}) admits the six parameter symmetry \cite{AKH00}
\begin{eqnarray} \label{e}
G &=& \theta(t) \partial_t + \Bigg( \gamma(t) + \frac{\log(S) \dot{\theta}(t)}{2} + \frac{\log(S) \dot{X}(t) \theta(t)}{2X(t)} \Bigg) S \partial_S + \Bigg( V(S,t) \alpha(t)
\nonumber \\ \noalign{\vskip 0.01cm}
&& \mbox{} + \Bigg( \dot{Z}(t)\log(S)\theta(t) + Z(t) \log(S) \dot{\theta}(t) - \log(S) \dot{\alpha}(t) + \Big[ Y(t) \log^2(S) - X(t)^2 \log^2(S) 
\nonumber \\ \noalign{\vskip 0.01cm}
&& \mbox{} - 2X(t)\log(S) \Big]k_1  \Bigg) V(S,t) \Bigg/ \Bigg( Y(t) - X(t) \Bigg) \Bigg) \partial_V,
\end{eqnarray}
subject to the functions $\theta(t), \alpha(t)$ and $\gamma(t)$  satisfying
\begin{align}
\label{b}
& X(t)^2 \ddot{\theta}(t) - 8X(t)k_1 - \dot{X}(t)^2\theta(t) + X(t) \dot{X}(t) \dot{\theta}(t) + X(t) \ddot{X}(t) \theta(t) = 0,\\
\label{c}
& \mbox{} -3X(t)^2\dot{Y}(t)\dot{\theta}(t) - 2X(t)^2\ddot{Y}(t)\theta(t) +  2X(t)^2 \ddot{\gamma}(t) + 3X(t) \dot{X}(t) Y(t) \dot{\theta}(t) + 2X(t)\ddot{X}(t)Y(t) \theta(t) 
\nonumber \\ \noalign{\vskip 0.01cm}
& \mbox{} + 3 X(t) \dot{X}(t) \dot{Y}(t) \theta(t) - 2 X(t) \dot{X}(t) \dot{\gamma}(t) - 3\dot{X}(t)^2 Y(t) \theta{t} = 0,\\
\label{d}
& 2X(t)^2 Y(t)\dot{\theta}(t) - 4X(t)^2 Z(t) \dot{\theta}(t) - X(t)^2 \dot{X}(t) \theta(t) + 2X(t)^2\dot{Y}(t) \theta(t) - 4X(t)^2 \dot{Z}(t)\theta(t) - 2X(t)^2\dot{\gamma}
\nonumber \\ \noalign{\vskip 0.01cm}
& \mbox{}  + 4X(t)^2 \dot{\alpha}(t) - X(t)^3 \dot{\theta}(t) + 8 X(t)^3 k_1 + \dot{X}(t) Y(t)^2 \theta(t) - X(t) Y(t)^2 \dot{\theta}(t) - 2 X(t) Y(t) \dot{Y}(t) \theta(t) 
\nonumber \\ \noalign{\vskip 0.01cm}
& \mbox{} + 2 X(t) Y(t) \dot{\gamma}(t) = 0.
\end{align}
Since the equation is linear, it also admits an infinite-dimensional `solution' symmetry \cite{OP86}.  The symmetry structure in (\ref{e}) and the accompanying equations in the system (\ref{b})--(\ref{d}) is applicable to all evolution-type equations of the form (\ref{a}). We will show its applicability by considering an example of exotic options.
\subsection{Solution to the power option problem} We now consider the power option that pays out a continuous yield dividend, $y(t)$, at different time intervals. To derive the PDE modeling this option, we assume that the stock pays a dividend  at a continuous rate proportional to the value of the stock. The dividend paid per unit time will thus be $y(t)S$ and the dividend paid in a short time interval will be $y(t)S\text{d}t$. The pricing kernel is characterized by the Brownian  motion model
\begin{equation}\label {1}
\frac{\text{d}\psi}{\psi} = -r(t) \text{d}t - \frac{\mu(t) - r(t)}{\sigma(t)} \text{d}W(t),
\end{equation}
 and the stochastic process for $S^\beta$ is given by
\begin{equation}\label{2}
\text{d}S^{\beta} = \Big[ (\mu(t) - y(t))\beta + \frac{1}{2}\beta (\beta - 1)\Big] S^\beta \text{d}t + \sigma(t) \beta S^\beta \text{d}W(t),
\end{equation}
where $r(t)$ is the stock interest rate, $\sigma(t)$ its volatility, $\mu(t)$ the instantaneous expected return and $W(t)$ is the Weiner process. By It\^{o}'s lemma, the price of the power derivative will be 
\begin{equation}\label{3}
\text{d}V = \Bigg( \frac{1}{2} \sigma(t)^2 \beta^2 S^{\beta 2} V_{S^{\beta} S^{\beta}} + V_t +  \Big[ (\mu(t) - y(t))\beta + \frac{1}{2}\beta (\beta - 1)\sigma(t)^2 \Big] S^\beta V_{S^\beta} \Bigg) \text{d}t + \sigma(t) \beta S^\beta V_{S^\beta} W(t).
\end{equation}
On solving for  $\psi$  and $V$ in equations (\ref{1}) and (\ref{3}) respectively, their product (since it is a martingale the drift term must be zero) gives the PDE
\begin{equation}\label{4}
V_t + \frac{1}{2} \sigma(t)^2 \beta^2 S^{2} V_{SS} +   \Big[ (r(t) - y(t))\beta + \frac{1}{2}\beta (\beta - 1)\sigma(t)^2 \Big] S V_{S} - r(t)V = 0,
\end{equation}
where the payoff is given by $V(S,T) = \text{max} \{\Psi(S^\beta - K),0 \}$  at the boundary $t = T$. The option has a payoff  $V(S,T) = 0$ when $S^\beta < K$ and a payoff $V(S,T) = S^\beta - K$ when $S^\beta \geq K$. The strike price is $K$ and $\Psi$ determines whether the option is a call or put. 

On substituting the representations of $X(t), Y(t)$ and $Z(t)$ of equation (\ref{4}) into the system (\ref{b}) -- (\ref{d}) and solving, we obtain
\begin{align}
\label{14}
& \theta(t) = \frac{1}{\sigma(t)^2} \Bigg[ \theta_2 + \int \sigma(t)^2 \text{d}t \Big( \theta_1 + 2k\beta^2 \int \sigma(t)^2 \text{d}t\Big) \Bigg],\\
\label{15}
%&& \gamma(t) =  \gamma_2 + \int \Bigg[\sigma(t)^2\Bigg(\gamma_1 + \beta \int \frac{\text{d}t}{\sigma(t)^2} \Bigg\{\Bigg[-\frac{3\dot{r}}{2} + \frac{3 r \dot{\sigma}(t)}{\sigma(t)} - \frac{3y(t)\dot{\sigma}(t) }{\sigma(t)} + 3\dot{r}(t)\dot{\sigma}(t)\Bigg]    
%\nonumber \\ \noalign{\vskip 0.01cm}
%&&  \Bigg[\theta_1 +   \int \sigma(t)^2 \text{d}t \Bigg( 4k\beta^2 - \frac{2\dot{\sigma}(t)}{\sigma(t)^3} \Bigg(\theta_1 + 2k\beta^2 \int \sigma(t)^2 \text{d}t  \Bigg) \Bigg) \Bigg] + \Bigg[  \int \sigma(t)^2 \text{d}t \Big( 2k\beta^2 \int \sigma(t)^2 \text{d}t
%\nonumber \\ \noalign{\vskip 0.01cm}
%&&  \theta_1\Big) + \theta_2 \Bigg] \Bigg[\frac{3\dot{y}(t) \dot{\sigma}(t)}{\sigma(t)} -  \frac{4\dot{r}(t) \dot{\sigma}(t)^2}{\sigma(t)^2} + \frac{4{y}(t) \dot{\sigma}(t)^2}{\sigma(t)^2} + \ddot{y}(t) + \frac{2{r}(t) \ddot{\sigma}(t)}{\sigma(t)} - \frac{2\dot{y}(t) \ddot{\sigma}(t)}{\sigma(t)} \Bigg]  \Bigg\}  \Bigg) \Bigg]\text{d}t \\
& \gamma(t) =  \gamma_2 + \int \Bigg[\sigma(t)^2\Bigg(\gamma_1 + \beta \int \frac{\text{d}t}{\sigma(t)^2} \Bigg\{\Bigg[-\frac{3\dot{r}}{2} + \frac{3 r \dot{\sigma}(t)}{\sigma(t)} - \frac{3y(t)\dot{\sigma}(t) }{\sigma(t)} + 3\dot{r}(t)\dot{\sigma}(t)\Bigg] \dot{\theta}(t)
\nonumber \\ \noalign{\vskip 0.01cm}
&\mbox{} + \Bigg[\frac{3\dot{y}(t) \dot{\sigma}(t)}{\sigma(t)} -  \frac{4\dot{r}(t) \dot{\sigma}(t)^2}{\sigma(t)^2} + \frac{4{y}(t) \dot{\sigma}(t)^2}{\sigma(t)^2} + \ddot{y}(t) + \frac{2{r}(t) \ddot{\sigma}(t)}{\sigma(t)} - \frac{2\dot{y}(t) \ddot{\sigma}(t)}{\sigma(t)} \Bigg] \theta(t) \Bigg\}  \Bigg) \Bigg]\text{d}t,\\
\label{16}
&\alpha(t) = \int \Bigg\{  \Bigg(r(t)\sigma(t)^3 - 2r(t)y(t)  + r(t)^2\sigma(t) + y(t)^2 \sigma(t) + y(t) \sigma(t)^3 + \frac{\sigma(t)^5}{4} \Bigg) \frac{\dot{\theta}(t)}{2\sigma(t)^3}
\nonumber \\ \noalign{\vskip 0.01cm}
&\mbox{} + \Bigg( y(t) + \frac{\sigma(t)^2}{2} - r(t)\bigg)\frac{\dot{\gamma}(t)}{\sigma(t)^2 \beta} - \sigma(t)^2 \beta^2 k_1 + \Bigg(\frac{\dot{r}(t) \sigma(t)^3}{2} - r(t)^2 \dot{\sigma}(t) - y(t)^2 \dot{\sigma}(t)
\nonumber \\ \noalign{\vskip 0.01cm}
&\mbox{} + 2r(t)y(t)\dot{\sigma}(t) + r(t)\dot{r}(t)\sigma(t) - r(t)\dot{y}(t)\sigma(t) + y(t) \dot{y}(t) \sigma(t) - y(t) \dot{r}(t) \sigma(t)  
\nonumber \\ \noalign{\vskip 0.01cm}
&\mbox{} + \frac{\dot{y}(t) \sigma(t)^3}{2} + \frac{\dot{\sigma(t)} \sigma(t)^4}{4} \Bigg)\theta(t) \Bigg\} \text{d} t + \alpha_1,
\end{align}
where $\theta_1, \theta_2, \gamma_1, \gamma_2$ and $\alpha_1$ are constants of integration.
\subsubsection{Solution to the power option problem with payoff $V(S,T) = 0$} Application of the general symmetry (\ref{e}) to the terminal condition will demand that $\theta(T) = 0$. For this to be true, the parameters $\theta_1, \theta_2$ and $k_1$ must be zero. This requirement reduces the 6-parameter symmetry to a 3-parameter symmetry. This new symmetry has invariants $P$ and $u$ resulting in 
\begin{equation}
V = P(u) \exp \Bigg(\frac{\log(S)}{\gamma(t) \beta}  \Bigg[\alpha_1 \beta - \frac{\gamma_1 \log(S)}{2\beta} + \gamma_1 \int \Bigg\{r(t)-y(t)-\frac{\sigma(t)^2}{2} \Bigg\}\text{d}t \Bigg]  \Bigg),
\end{equation}
where $u = t$. We use these to reduce the power option PDE to an ODE with solution
\begin{equation}
P = C_1  \Bigg( \cosh \Big( {u A_1(t)}\Big/{A_2(t)} \Big) - \sinh \Big( {u A_1(t)}\Big/{A_2(t)} \Big)\Bigg) %\exp \Big( {-u A_1(t)}\Big/{A_2(t)} \Big)
\end{equation}
where the $A_i$'s are known functions of time. The solution of the PDE satisfying the terminal condition will thus be 
\begin{equation}
V = C_1 \Bigg( \cosh \Big( {u A_1(t)}\Big/{A_2(t)} \Big) - \sinh \Big( {u A_1(t)}\Big/{A_2(t)} \Big)\Bigg) S^{ \frac{1}{\gamma(t) \beta}  \Bigg[\alpha_1 \beta - \frac{\gamma_1 \log(S)}{2\beta} + \gamma_1 \int \Bigg\{r(t)-y(t)-\frac{\sigma(t)^2}{2} \Bigg\}\text{d}t \Bigg] } 
%V(S,t) = C_1 \exp \Bigg(-\frac{t A_1(t)}{A_2(t)}\Bigg) S^{ \displaystyle \frac{1}{\gamma(t) \beta}  \Bigg[\alpha_1 \beta - \frac{\gamma_1 \log(S)}{2\beta} + \gamma_1 \int %\Bigg\{r(t)-y(t)-\frac{\sigma(t)^2}{2} \Bigg\}\text{d}t \Bigg] } 
\end{equation}
where the constant $C_1$ is determined by the requirement that $V(S,T)=0$.
\subsubsection{Solution to the power option problem with payoff $V(S,T) = S^\beta - K$} 
For the condition $V(S,T) = S^\beta - K$ at $t = T$, we follow the same steps as for the terminal condition. The general symmetry in equation (\ref{e}) reduces to a one-parameter symmetry with the constants $\theta_1, \gamma_1, \gamma_2, \alpha_1$ and $k_1$ all equal to zero. The function $\theta(t)$ is redefined as
\begin{equation}
\theta^1(t) = \frac{\theta_2}{\sigma(t)^2} \Bigg(1 - \int \sigma(t)^2 \text{d}t \Bigg|_{t=T} \int \sigma(t)^2 \text{d}t \Bigg),
\end{equation}
%Same steps as for the terminal condition will then be followed. 
with obvious modifications for $\gamma^1(t)$ and $\alpha^1(t)$ (see (\ref{15}) and (\ref{16})). The solution will then have the form
\begin{equation}
V = P^1(u) \exp\Bigg( \int \Bigg\{ \frac{\alpha^1(t)}{\theta^1(t)} - \frac{2\log(S)}{2r(t)\beta - \beta \sigma(t)^2 - 2 \beta y(t)} \Bigg[ \frac{\dot{\alpha}^1(t)}{\theta^1(t)} - \frac{r(t) \dot{\theta}^1(t)}{\theta^1(t)}  - \dot{r}(t)\Bigg]\Bigg\}  \text{d}t \Bigg) ,
\end{equation}
where 
\begin{equation}
u = \int \frac{\gamma^1(t)}{\sigma(t) (\theta^1(t))^2}\text{d}t - \frac{\log(S)}{\sigma(t) \theta^1(t)}.
\end{equation}
These reduce the power option PDE to a second order ODE with solution 
\begin{align}
P^1 =& \Bigg[ \cosh \Bigg(2a(t) + ub(t) - \frac{2c(t)}{b(t)}\Bigg) - \sinh \Bigg(2a(t) + ub(t) - \frac{2c(t)}{b(t)}\Bigg) \Bigg] \Bigg[ C_1 H\Bigg(-1+\frac{c(t)^2}{b(t)^3} +\frac{a(t) c(t)}{b(t)^2};
\nonumber \\ \noalign{\vskip 0.01cm}
& \mbox{}\frac{a(t)b(t) + ub(t)^2 -2c(t)}{\sqrt{2b(t)^3}} + C_2\hspace{0.05cm} {}_1F_1 \Bigg(\frac{b(t)^3 +a(t)b(t)c(t) - c(t)^2}{2b(t)^3};\frac{1}{2}; \frac{\Big( a(t)b(t) +ub(t)^2 -2c(t)^2\Big)^2 }{2b(t)^3} \Bigg) \Bigg],
\end{align}
where $H$ and ${}_1F_1$ are the Hermite and Hypergeometric functions respectively. As before $a(t), b(t)$ and $c(t)$ are known functions of time. The solution to the problem with payoff $V(S,T) = S^\beta - K$ will thus be
\begin{align}
V(S,T) =& \exp\Bigg[ \int \frac{\alpha(t)}{\theta(t)} \text{d}t  -\frac{u}{2} \Bigg(2a(t) + ub(t) - \frac{2c(t)}{b(t)}\Bigg)  \Bigg] \Bigg[ C_1 H\Bigg(-1+\frac{c(t)^2}{b(t)^3} +\frac{a(t) c(t)}{b(t)^2};
\nonumber \\ \noalign{\vskip 0.01cm}
& \mbox{}\frac{a(t)b(t) + ub(t)^2 -2c(t)}{\sqrt{2b(t)^3}}\Bigg) + C_2\hspace{0.05cm} {}_1F_1 \Bigg(\frac{b(t)^3 +a(t)b(t)c(t) - c(t)^2}{2b(t)^3};\frac{1}{2}; \frac{\Big( a(t)b(t) +ub(t)^2 -2c(t)^2\Big)^2 }{2b(t)^3} \Bigg) \Bigg]
\nonumber \\ \noalign{\vskip 0.01cm}
& S^ {\displaystyle 2\int \frac{\dot{\alpha}(t) - r(t)\dot{\theta}(t) - \dot{r}(t)\theta(t) }{\theta\big( 2r(t)\beta - \beta \sigma(t)^2 - 2 \beta y(t) \big)} \text{d}t}.
\end{align}
The constants $C_1, C_2$ are determined from the requirement that $V(S,T) = S^\beta - K$.
\section{Conclusion and remarks} 
We have shown that all evolution equations of the form (\ref{a}), regardless of the nature of the coefficients, will have the 6-parameter symmetry structure (\ref{e}) whose solution route will follow the same steps outlined in this paper. As a result, we can find group invariant solutions to a large class of financially relevant DEs.  

We introduced  a new PDE modelling power options with time dependent parameters and paying out a continuous dividend at different time intervals. Our general result was then contextualized to this equation. New solutions satisfying the boundary conditions were provided for the first time. 

We believe that this approach will be useful in Financial Mathematics.
  
%The explicit nature of the solutions lend themselves to more effective modeling than numerical solutions.

{\bf Acknowledgements:} We thank the University of KwaZulu-Natal and the National Research Foundation for their ongoing support. 

\small{
}

\begin{thebibliography}{99}
%\textbf{Bibliography}
\bibitem{BS73}
Black, F. and Scholes, M.: The pricing of options and corporate liabilities, J. Pol. Econ., 81 637–654 (1973)
\bibitem{BAA08}
 Bodie, Z, Alex K. and Alan J. M., Investments (7th ed.), McGraw-Hill/Irwin, New York 2008
\bibitem{CR78}
J. Cox, M. Rubinstein, Option Markets, Prentice-Hall, New Jersey, 1978
\bibitem{TRG99}
R.G. Tompkins, Power options: hedging nonlinear risks, J. Risk 2 (1999) 29--45
\bibitem{ESS04}
A. Esser, Pricing in (in)Complete Markets: Structural Analysis and Applications, Springer-Verlag, New York 2004
\bibitem{OGO14}
Okelola M. O., Govinder K. S. and O'Hara J. G., Solving a PDE associated with the pricing of power options with time dependent parameters, Accepted and about to appear in Mathematical methods in the Applied Sciences (2014)
\bibitem{OP86}
Olver, P.: Applications of Lie groups to differential equations, Springer-Verlag, New York 1986
\bibitem{JM01}
I.K. Johnpillai, F.M. Mahomed, Singular invariant equation for the (1+1) Fokker-Planck equation, J. Phys. A: Math. Gen. 34 (2001) 11033--11051 
\bibitem{COG10}
Caister, N. C., O'Hara, J. G. and Govinder, K. S.: Solving the Asian option PDE using Lie symmetry methods, Intl. J. Theor. App. Fin., 13 1265–1277 (2010)
\bibitem{LOS06}
P. G. L. Leach, O'Hara J. G. and Sinkala W.,  Symmetry-based solution of a model for a combination of a risky investment and a riskless investment, J. Math. Anal. App., DOI: 10.1016/j.jmaa.2006.11.056
\bibitem{AKH00}
A.K. Head, LIE, a PC program for Lie analysis of differential equations, Comp. Phys. Comm.  71 (1993) 241--248
\end{thebibliography}
\end{document}